\documentclass[12pt]{article}
\usepackage{epsfig}

\def\beq{\begin{equation}}
\def\eeq#1{\label{#1}\end{equation}}
\def\eeqn{\end{equation}}

\def\beqa{\begin{eqnarray}}
\def\eeqa#1{\label{#1}\end{eqnarray}}
\def\eeqan{\end{eqnarray}}

\let\bar=\overbar

\def\Dslash{\not{\hbox{\kern-4pt $D$}}}
\def\dslash{\not{\hbox{\kern-2pt $\del$}}}

\def\msb{{\bar{\ssstyle M \kern -1pt S}}}

\usepackage{fancyhdr,graphicx}
\def\Title#1{\begin{center} {\Large {\bf #1} } \end{center}}

\begin{document}

\Title{\baselineskip=10pt
The pasta structure in the hadron-quark\\
phase transition and the effects on\\
magnetized compact stars\\
}

\bigskip\bigskip


\begin{raggedright}

{\it Nobutoshi Yasutake\index{Yasutake, N.} $^{1*}$,
Toshiki Maruyama\index{Maruyama, T.} $^2$,
Toshitaka Tatsumi\index{Tatsumi, T.} $^3$,\\
Kenta Kiuchi\index{Kiuchi, K.} $^4$ and
Kei Kotake\index{Kotake, K.} $^1$\\
\bigskip

$^1$ Division of Theoretical Astronomy,
National Astronomical Observatory of Japan,
2-21-1 Osawa, Mitaka, Tokyo 181-8588,
Japan \\
\medskip

$^2$ Advanced Science Research Center,
Japan Atomic Energy Agency,
Tokai, Ibaraki 319-1195,
Japan \\
\medskip

$^3$ Department of Physics, Kyoto University,
Kyoto 606-8502, 
Japan \\
\medskip

$^4$ Sience and Engineering, Waseda University,
3-4-1 Okubo,Shinjuku, Tokyo 169-8555, 
Japan \\
\medskip

{* \tt Email: yasutake@th.nao.ac.jp}
}
\bigskip\bigskip
\end{raggedright}

\section{Introduction}
So far, there has been extensive works devoted to studying the effects of quark matter on astrophysical phenomena; the gravitational wave radiations~\cite{lin06, yasutake07, abdikamalov08},  cooling processes~\cite{page00,blaschke00,blaschke01,grigorian05}, neutrino emissions~\cite{nakazato08, sagert08}, rotational frequencies~\cite{burgio03}, the maximum energy release by conversions from neutron stars to quark/hybrid stars~\cite{yasutake05, zdunik07}, etc.. However, uncertainties of equation of state~(EOS) have been still left.

For such studies on compact stars, general relativistic effects are
fundamentally important, since baryon density and strong magnetic field are comparable to pressure,
$\rho_0 c^2 \sim B^2/8\pi \sim P$. 
Here we report the effects of quark-hadron phase transition on the structures of general relativistic stars with purely toroidal magnetic field.
For the mixed phase, we take into account of the finite-size effects, which lead to  non-uniform ``Pasta" structures.
The star with pure toroidal magnetic field is unstable,
but it becomes another stable star in which toroidal magnetic field is dominant in the dynamical simulation~\cite{kiuchi08b}. Moreover, Heger et al. have suggest that the toroidal magnetic field may dominate $10^5$ times lager than the poloidal magnetic field at the last stage of the main sequence~\cite{heger05}. 

The organization of this report is as follows. Adopted EOS is briefly discussed in Sec.~2 with  equilibrium models of magnetized rotating compact
stars. In
Sec.~3, we show our numerical results. In Sec.~4, we discuss the consequence
of our calculations. 

\section{Framework}
\subsection{Equation of state}
The EOS used in this study has been calculated in the previous study by Maruyama et al.\cite{maruyama07}.
The theoretical framework for the hadronic phase of matter is the
nonrelativistic Brueckner-Hartree-Fock approach including hyperons such as
$\Sigma^-$ and $\Lambda$~\cite{baldo98}.
There is a controversy about the $\Sigma^-$-$N$ interaction. 
The recent experimental result about hypernuclei has suggested that it is repulsive~\cite{noumi02,saha04}, 
while we use a weak but attractive interaction 
with which $\Sigma^-$ appears before $\Lambda$ in nuclear matter. 
It would be interesting to see how our results are changed 
by using different $\Sigma^-$-$N$ interaction
and we will discuss it in the future work.

For the quark phase, we adopt the MIT bag model consisting of
$u,~d~,s$-quarks.
Probably, this model is too simple to describe quark matter.
We will adopt more sophisticated model in the future~\cite{yasutake09a}.
We assume massless $u$ and $d$ quarks, and $s$-quark with a current mass of $m_s=180$ MeV. 
We set the bag constant $B$ to be 100~MeV fm$^{-3}$ in this report. 

We employ the Thomas-Fermi approximation for 
the   kinetic energies of hadrons and quarks.
For the mixed phase, we 
assume various geometrical structures of matter shown in Fig.\ 1. 
Between two phases we put a sharp boundary with  a constant surface tension.
We impose the Gibbs conditions, i.e.\ chemical equilibrium among particles in two phases 
consistent with the Coulomb potential, 
and a pressure balance consistent with the surface tension.
Unfortunately, there is a wide range of uncertainties about the 
surface tension, $\sigma_S \sim 10$--100 MeV~fm$^{-2}$~\cite{farhi84,huang90,kajantie91}. 
In this study, we use two values $\sigma_S=10$ and 40 MeV~fm$^{-2}$.

\begin{figure}[tb]
\begin{center}
\epsfig{file=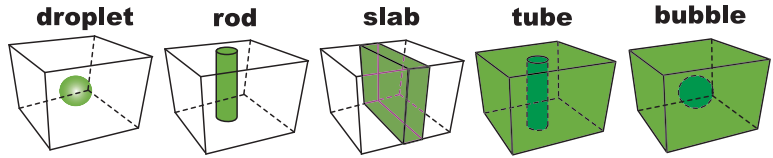,width=4in}
\caption{
Schematic picture of Pasta structures, where matter in one phase is immersed in another phase in phase quilibrium.  
We assume that one of these structures appears 
in a Wigner-Seitz cell with a geometrical symmetry.
}
\label{fig:pasta}
\end{center}
\end{figure}
 
\begin{figure}[hbt]
\begin{center}
\includegraphics[width=75mm]{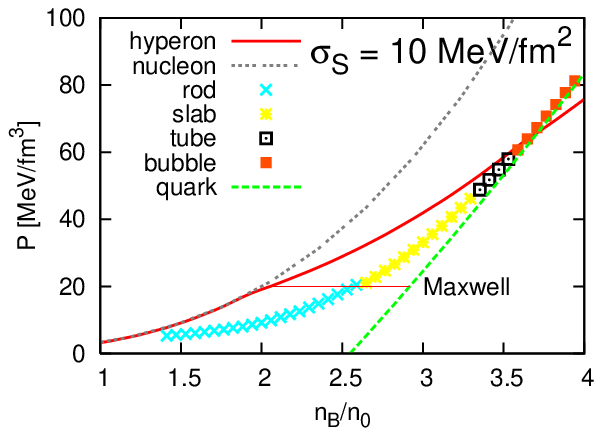} 
\includegraphics[width=75mm]{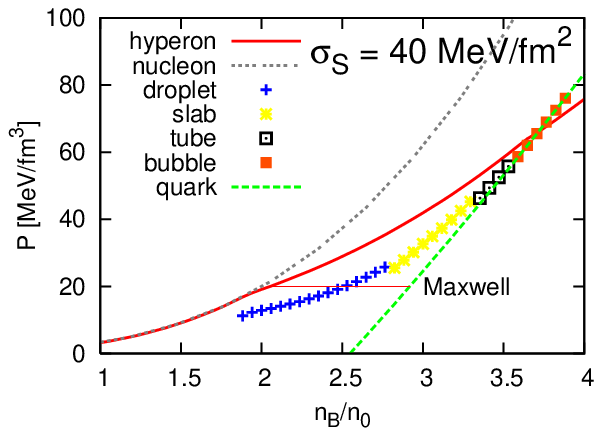} 
\caption{\label{fig:eos_mix} Pressure versus baryon density for each matter. The left~(right) panel shows EOS in the case of $\sigma_S$ = 10(40) MeV~fm$^{-2}$. Each thick symbols show the mixed phase with different geometrical structures as shown  in Fig.~1. Solid~(dashed) line shows a pure hadron~(quark) matter phase. Dotted line shows the hadron phase without hyperons. For comparison, we also shows the Maxwell construction case. }
\end{center}
\end{figure}

Now we show our EOS in Fig.~\ref{fig:eos_mix}. 
We impose the barotropic condition of the EOS ($P=P(\epsilon)$) by assuming zero-temperature, neutrino-free, and beta-equilibrium matter. 
Two panels show the pressure versus baryon density for the quark-hadron mixed phase. For the weak surface tension~($\sigma_S = 10$ MeV~fm$^{-2}$), the droplet structure does not appear, wheres, for the strong surface tension~($\sigma_S = 40$ MeV~fm$^{-2}$), the rod structure does not appear. Furthermore,  the mixed phase is restricted and EOS gets close to that by the Maxwell construction  for the strong surface tension,.

\subsection{Equilibrium models of magnetized rotating compact stars}
Master equations for the rotating relativistic stars containing purely toroidal magnetic fields are based on the assumptions summarized as follows~\cite{kiuchi08,yasutake09b}; (1) Equilibrium models are stationary and axisymmetric. (2) The matter source is approximated by a perfect fluid with infinite conductivity. (3) There is no meridional flow of matter. (4) EOS for matter is barotropic. (5) Magnetic axis and rotation axis are aligned. This barotropic condition can be satisfied for our EOS. 

\section{Numerical Results}\label{sec:result}
Now, we show the configurations of strongly magnetized compact stars
with/without the quark-hadron mixed phase. 
In this report, we consider non-rotating static configurations since magnetars observed so far are all slow rotators.
In Fig.~3, we show distributions of the baryon density and the magnetic field in the
meridional planes for the static-equilibrium stars characterized 
by (1)
$B_{\rm max}=6.2 \times 10^{17}$~G, $M=1.30M_\odot$ with the quark-hadron phase transition~($\sigma
=10$MeV/fm$^2$) [the first panel]
, (2)
$B_{\rm max}=6.2 \times 10^{17}$~G, $M=1.31M_\odot$ with the quark-hadron phase transition~($\sigma
=40$MeV/fm$^2$) [the second panel]
, (3)
$B_{\rm max}=7.1 \times 10^{17}$~G, $M=1.31M_\odot$ with the hyperon EOS [the third panel]
, and by (4) 
$B_{\rm max}=6.2 \times 10^{17}$~G, $M=1.32M_\odot$ with the nucleon EOS [the last panel]. 
The meaning of each physical quantity is as follows; the maximum strength of the magnetic field for 
$B_{\rm max}$ and the gravitational mass for $M$. 
All the models have the same magnetic flux of $5.00\times10^{29}{\rm G~cm^2}$ and the same baryon mass of $1.45 M_\odot$. 

\begin{figure}[hbt]
\begin{center}
\includegraphics[width=11cm]{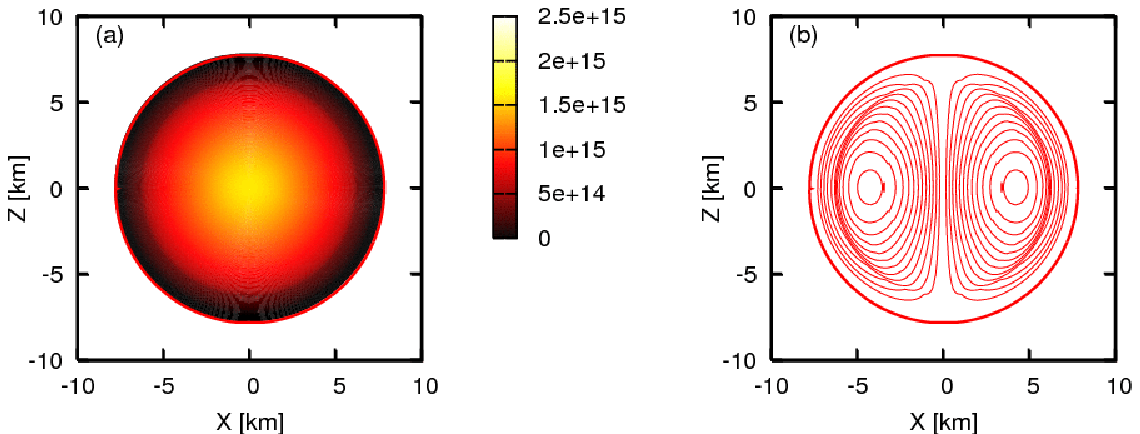} \\
\includegraphics[width=11cm]{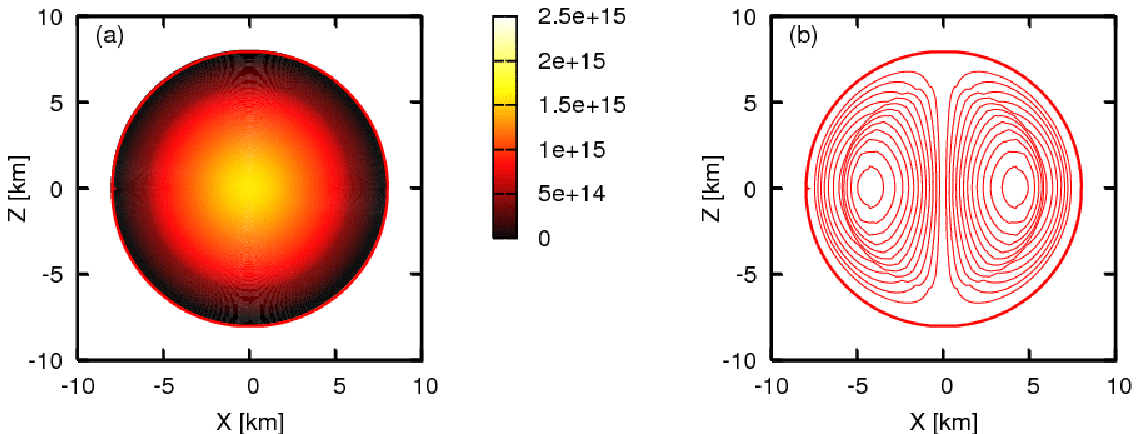} \\
\includegraphics[width=11cm]{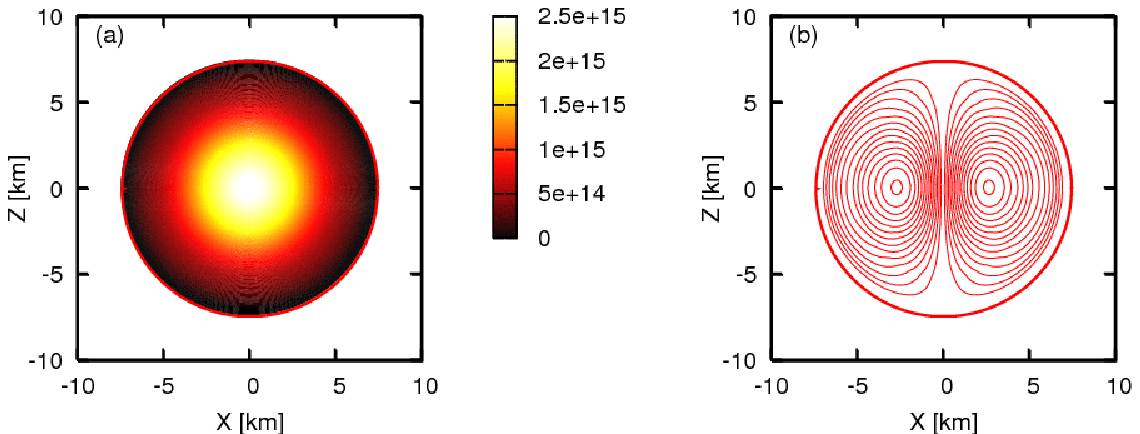}       \\
\includegraphics[width=11cm]{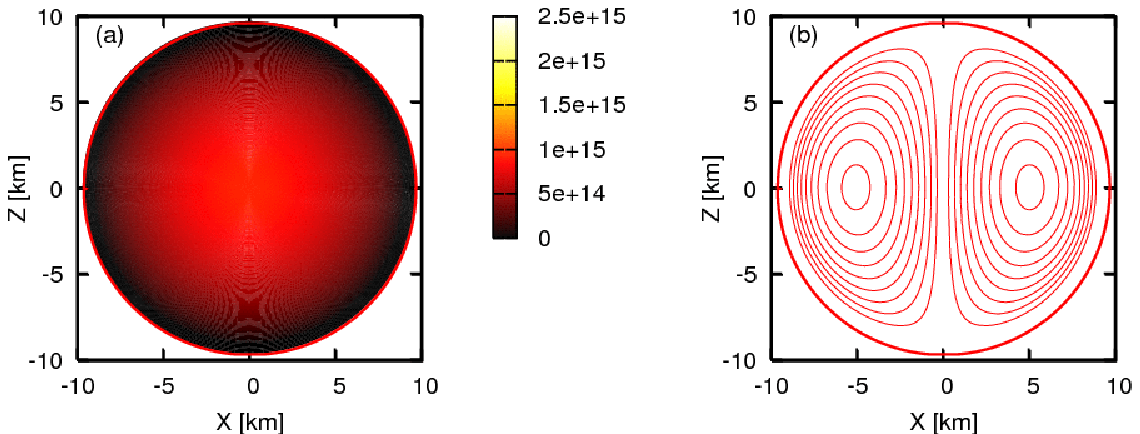} 
\caption{\label{fig:dis_mix} Distributions of (a): rest mass density [g/cm$^3$] and (b): magnetic field [G] with same magnetic flux and the baryon mass.
EOSs of the panels are EOS with the phase transition of $\sigma=10$ MeV/fm$^{-2}$[the first panel], 
 $\sigma=40$ MeV/fm$^{-2}$[the second panel], the hyperon EOS~[the third panel], 
and the nucleon EOS~[the last panel]. 
}
\end{center}
\end{figure}

Clearly, the distribution of magnetic field is different between two cases. The toroidal magnetic field lines behave like a rubber belt wrapped around the waist of the stars with the hadronic EOS (see the lower two panels of Fig.~3). However, for a hybrid star, the distribution of magnetic field has a discontinuity for the equatorial direction for any value of $\sigma_S$~(see the upper two panels of Fig.~\ref{fig:dis_mix}). We can understand this easily; The magnetic field is frozen in matter, so that the distribution of magnetic field depends on the distribution of density. Hybrid stars have discontinuities in the density profiles due to the phase transition; e.g. in Fig.~\ref{fig:dis_mix}, the cores~(inner 6 km) of hybrid stars are the mixed-phase matter, and the baryon density in this density regions has a discontinuity, which then gives rise to the discontinuity of magnetic field as shown in Fig.~2. Such discontinuity of magnetic field will change the thermal conductivities of compact stars and change the cooling processes.

\section{Summary and Discussion}\label{sec:summary}
In this report, we have investigated the effects of the quark-hadron mixed phase on the magnetized rotating stars with the general relativistic equilibrium configuration. As a result, we find that the distribution of the magnetic field for hybrid stars has a discontinuity in the quark-hadron mixed phase. 

It was pointed out that the strong magnetic field may change EOSs~\cite{broderick00}. In particular for quark matter, it has been found that the energy gaps in the magnetic Color-Flavor-Locked phase are the oscillating functions with respect to the magnetic field~\cite{noronha07, fukushima08}. All of their effects are important, and may change our results. 

\bigskip

We thank K.~Kashiwa, M.~Hashimoto, S.~Yamada, S.~Yoshida, and Y.~Eriguchi for informative discussions.
This study was supported in part by the Grants-in-Aid for the Scientific Research from the Ministry of Education, Science and Culture of Japan (No. 20540267, 21105512, 19540313).






\end{document}